\documentclass[aps,pre,showpacs,twocolumn,superscriptaddress]{revtex4-1}
\usepackage{graphicx}
\usepackage{txfonts}
\usepackage{amssymb}
\usepackage{makeidx}

\def\Tr{\mathrm{Tr}}

\def\dd{\mathrm{d}}

\def\simge{\mathrel{%
    \rlap{\raise 0.511ex \hbox{$>$}}{\lower 0.511ex \hbox{$\sim$}}}}
\def\simle{\mathrel{
    \rlap{\raise 0.511ex \hbox{$<$}}{\lower 0.511ex \hbox{$\sim$}}}}

\newcommand \be{\begin{eqnarray}}
\newcommand \ee{\end{eqnarray}}
\newcommand{\del}{\partial}

\def\Xint#1{\mathchoice
{\XXint\displaystyle\textstyle{#1}}%
{\XXint\textstyle\scriptstyle{#1}}%
{\XXint\scriptstyle\scriptscriptstyle{#1}}%
{\XXint\scriptscriptstyle\scriptscriptstyle{#1}}%
\!\int}
\def\XXint#1#2#3{{\setbox0=\hbox{$#1{#2#3}{\int}$}
\vcenter{\hbox{$#2#3$}}\kern-.5\wd0}}

\def\dashint{\Xint-}

\begin{document}

\title{Universal shocks in the Wishart random matrix ensemble - 1}

\author{Jean-Paul Blaizot}
\email{Jean-Paul.Blaizot@cea.fr} \affiliation{IPTh,  CNRS/URA 2306, CEA-Saclay,
91191 Gif-sur Yvette, France}

\author{Maciej A. Nowak}
\email{nowak@th.if.uj.edu.pl} \affiliation{M. Smoluchowski Institute
of Physics and Mark Kac Center for Complex Systems Research,
Jagiellonian University, PL--30--059 Cracow, Poland}

\author{Piotr Warcho\l{}} \email{piotr.warchol@uj.edu.pl} \affiliation{M.
Smoluchowski Institute of Physics,  Jagiellonian University,
PL--30--059 Cracow, Poland}\affiliation{IPTh, CEA-Saclay,
91191 Gif-sur Yvette, France}
\date{\today}

\begin{abstract}  
We show that the  derivative of the logarithm of the  average characteristic polynomial of a diffusing Wishart matrix obeys an exact partial differential equation valid for an arbitrary value of $N$, the size of the matrix.  In the large $N$ limit, this equation generalizes the simple Burgers equation that has been obtained earlier  for Hermitian or unitary matrices. The solution through the method of characteristics presents singularities that we relate to the precursors of shock formation in  fluid dynamical equations. The $1/N$ corrections may be viewed as viscous corrections, with the role of the viscosity being played by the inverse of the doubled dimension of the matrix. These corrections are studied through a scaling analysis in the vicinity of the shocks, and one recovers in a simple way the universal Bessel oscillations  (so-called
hard  edge singularities) familiar in random matrix theory.  
\end{abstract}

\maketitle

\section{Introduction}

Recently, two of us have argued~\cite{BN1,BN2} that  Dyson's original  idea~\cite{DYSON} to interpret  the distribution of eigenvalues of a random matrix as the equilibrium limit of a random walk performed independently by each of the matrix elements, can be exploited in order to get a new insight on some properties of the spectrum of  eigenvalues. We have considered the two simplest, albeit non-trivial examples: additive and multiplicative matrix-valued Brownian motions. In both cases, the evolving spectra are governed by complex, non-linear diffusion equations ~\cite{BN1,BN2,NEUB1,NEUB2}. The solutions of these equations, based on the method of complex characteristics, exhibit singularities that we may associate to the precursors of shocks (which we refer to as `pre-shocks' ~\cite{CHKD}).  Interestingly, for finite values of $N$, corresponding to a non-zero ``spectral viscosity'' $\nu_s=\frac{1}{2N}$, these shocks are accompanied   by  universal, oscillatory phenomena. In the case of additive Gaussian random walk, the oscillations are those of an Airy function. 
In the case of the multiplicative matrix valued-diffusion ~\cite{JW}, the corresponding evolving matrices are unitary, and an additional phenomenon occurs when, due to the compactness of the support of eigenvalues (unitary circle), the left and right Airy shocks merge into a more complicated structure described by a Pearcey function. This universal dynamics has been observed, for example, in the behavior of Wilson loops in simulations of  Yang-Mills theory with a large number of colors~\cite{NN}. In this case,  the role of the time  is played by the area of the Wilson loop and  the size of the matrix is determined by the number of colors. The critical oscillations of the preshock  are associated to the onset of non-perturbative effects  of the Wilson loop, i.e. they signal the transition to the strong coupling regime.

In this paper,  we further develop this fluid dynamical perspective for the family of random matrices known as the Wishart ensemble\cite{WISHART}. Originally formulated as the multivariate analogue of the gamma distribution, the Wishart ensemble, plays, till today, a dominant role in multivariate statistical data analysis, including such modern fields as image processing, genetic chips, financial engineering, wireless telecommunication and several others~\cite{MODERNWISHART}. Complex Wishart ensembles are also closely related to chiral random matrix models, defining the universal regime of spectral properties of Dirac operators in Euclidean non-abelian gauge theories ~\cite{VERBAARSCHOTREVIEW}. 

We start by  formulating the random walk of complex Wishart ensembles in terms of a  Smoluchowski-Fokker-Planck equation for the probability distribution of eigenvalues. From this equation, after proper rescaling of time, and using standard techniques of statistical mechanics, we recover in the large $N$ limit, a complex non-linear equation of the Burgers type for the resolvent (whose imaginary part yields the average spectral density).  We solve this equation by the method of complex characteristics, and recover the well-known spectral properties of the ensemble (Bronk-Marcenko-Pastur equation)~\cite{BRONK,MARPAS}. We also make a link to the Voiculescu~\cite{VOICULESCU} free random variables theory. 
We proceed then to the analysis of finite $N$ effects in order to capture the  universal behavior that is anticipated in the vicinity of the singular points associated with the edges of the spectrum and that correspond to  pre-shocks in the Burgers equation. In order to do so, we proceed as  in the prequels of this paper~\cite{BN2,NEUB1}, and focus on the time-dependent average characteristic polynomial, more precisely the derivative of its logarithm, which, in the large $N$ limit satisfies the same Burgers equation as the resolvent. An exact  nonlinear partial differential equation can be obtained for this object, that is valid for arbitrary $N$, and  that therefore allows us to study the oscillatory behavior in vicinity of the shocks that is visible for large but finite $N$. In particular, at one of the edges, we obtain the Bessel equation, whose  solution defines the universal behavior of the spectral functions near the origin of the spectrum, i.e. so-called ``hard edge" singularity. We conclude the paper with the study of the evolution of inverse characteristic polynomials. The Cauchy transforms of the associated monic orthogonal polynomials fulfill the same nonlinear Burgers-like partial differential equations.
\section{The random walk of Wishart matrices}
We consider a $N\times N$ random matrix of the form $L=K^{\dagger}(t)K(t)$ belonging to the complex ($\beta=2$) Wishart ensemble. The matrix $K$ is rectangular $M\times N$ ($M>N$) and with complex entries. We assume that the elements of  $K$ evolve in `time' according to ${\rm d}K_{ij}(t)=b^{(1)}_{ij}(t)+ib^{(2)}_{ij}(t)$, where $b^{(1)}_{ij}(t),b^{(2)}_{ij}(t)$ are real numbers defined by two independent  Brownian walks: \be b^{(c)}_{ij}(t)=\zeta^{(c)}_{ij}(t)\,{\rm d}t,\label{random walk}\ee with \be\left\langle \zeta^{(c)}_{ij}(t)\right\rangle = 0\ee and \be\left\langle \zeta^{(c)}_{ij}(t)\zeta^{(c')}_{kl}(t')\right\rangle =\delta^{cc'}\delta^{ik}\delta^{jl}\delta(t-t').\ee This is sometimes referred to as a Laguerre process, an analogue of the non-colliding squared Bessel process \cite{KONIG}. The real valued equivalent ($b^{(2)}_{ij}(t)=0$, $\beta=1$) is called the Wishart process,  and was first studied in ~\cite{BRU, BRU2}.   

Following Dyson \cite{DYSON}, we shall study the behavior of the  eigenvalues of $L$ as a function of time. Note however that since we are not interested in the equilibrium distribution, we do not include any restoring force on the motion of the eigenvalues. That is, the dynamics is entirely given by the random walks described above.  The random matrix $K$ admits a singular value decomposition $K=U\kappa V$, with $U,V$ unitary, and $\kappa={\rm diag}(\kappa_1,\cdots,\kappa_N)$ rectangular. The eigenvalues of $L$, which we call $\lambda_{i}$, are related to the non-singular values of $K$ by $\lambda_i=\kappa_i^2$. Furthermore, it is worth mentioning, that the pairs $-\kappa_i$ and $\kappa_i$ are the nonzero eigenvalues of the complex chiral matrix:
\[W=\left(\begin{array}{cc} 0 & K^{\dagger} \\ K & 0 \end{array}\right).\] 
Note, moreover, that the $M\times M$ (`anti-Wishart') matrix  $L_{a}=K K^\dagger$ has the same set of non vanishing eigenvalues as the $N\times N$ matrix $L$. In addition, both $L_{a}$ and $W$ possess $M-N\equiv\nu$ zero eigenvalues. 
The easiest (and physically most transparent) way to obtain the time evolution of the $\kappa_i$'s,  is to use second order perturbation theory \cite{DYSON}. A simple generalization of the calculation presented  in Ref.~\cite{AKUWAD} yields:
\be\nonumber
&&\langle \delta\kappa_i\rangle= \left\{  \frac{\beta(\nu+1)-1}{2\kappa_i}+\frac{\beta}{2}     \sum_{j(\neq i)}\left[\frac{1}{\kappa_i-\kappa_j}+\frac{1}{\kappa_i+\kappa_j}\right]\right\} \delta t, \nonumber\\&& \langle   \delta\kappa_i\delta\kappa_j \rangle=\delta_{ij}\,\delta t.\label{stokap}
\ee
To get the evolution of the $\lambda_i$'s, note that
\be
\lambda_i'=(\kappa_i+\delta\kappa_i)^2=\lambda_i+2\kappa_i \delta\kappa_i+\delta\kappa_i^2,
\ee
so that, upon averaging,  (with $\delta\lambda_i=\lambda_i'-\lambda_i$)
\be
\langle \delta\lambda_i\rangle=2 \kappa_i \langle \delta \kappa_i \rangle+\langle \delta\kappa_i^2\rangle.
\ee
Using this relation and  Eq.~(\ref{stokap}) one then easily obtains
\be\nonumber
&&\langle \delta\lambda_i\rangle=\beta\left[  \nu+1+2\lambda_i \sum_{j(\neq i)}
\frac{1}{\lambda_i-\lambda_j}\right]\, \delta t,\nonumber\\
&&\langle\delta \lambda_i\delta\lambda_j\rangle=4\lambda_i\,\delta_{ij} \,\delta t.
\label{stolam}\ee

Eqs.~(\ref{stokap}) and (\ref{stolam}) can be also derived via Ito calculus \cite{KATORI} and written in the following stochastic differential forms:
\be {\rm d}\kappa_i=b_i(t) +\frac{\beta}{2} \left\{\frac{\nu \!+\!1\!-\!\frac{1}{\beta}}{\kappa_{i}} +\sum_{j(\neq i)}\left[\frac{1}{\kappa_i\!-\!\kappa_j}+\frac{1}{\kappa_i\!+\!\kappa_j}\right]\right\}{\rm d}t \label{chiralLangevin}, \ee 
and
\be {\rm d}\lambda_i=2\sqrt{\lambda_i}
b_i(t) +\beta \left(\nu+1 +2\lambda_i\sum_{j(\neq i)}
\frac{1}{\lambda_i-\lambda_j}\right)
{\rm d}t. \label{complexbru} \ee 
 The latter can be interpreted as an equation for the evolution of $N$, non-intersecting, aka "vicious", walkers confined to the positive part of the real axis. It has a physical realization as, for example, a model for nonintersecting fluctuating interfaces in thermal equilibrium ~\cite{NADMAJ}. Recently, similar diffusion processes for the Wishart ensembles were also generalized for arbitrary value of parameter $\beta  \in (0, 2]$\cite{VIVO}.

All the statistical properties of the eigenvalues can in principle be derived from the probability  $P( \lambda_1,\lambda_2,\cdots, \lambda_N ,t)$  to find the eigenvalues near the values $\lambda_1,\cdots,\lambda_N$ at time $t$.  By using Eq.~(\ref{stolam}) or Eq.~(\ref{complexbru}), we may write a Smoluchowski-Fokker-Planck equation for $P$. It takes the following form: 
\be\nonumber
\frac{\del P}{\del t}&=&2\sum_{i}\frac{\del^{2}}{\del \lambda_{i}^{2}}\left(\lambda_{i}P\right)+\nonumber\\
&-&\beta\sum_{i}\frac{\del}{\del\lambda_{i}}\left[\left(\nu+1+2\lambda_{i}\sum_{j\left(\neq i\right)}\frac{1}{\lambda_{i}-\lambda_{j}}\right)P\right]. \label{SFP3}
\ee  
In this paper, we shall consider the specific case where the random walk (\ref{random walk}) is initiated at the origin, i.e., 
K$_{ij}(0)=0$ for all $i$'s and $j$'s.
In this case, it is easy to verify that the solution of Eq.~(\ref{SFP3}) is 
\be P=\mathcal{N}t^{-\frac{\beta}{2}MN}\prod_{i<j}(\lambda_j-\lambda_i)^{\beta}\prod_{k=1}^{N}\lambda_k^{\frac{\beta}{2}(\nu+1)-1}{\rm e}^{-\sum_{n=1}^{N}{\lambda_{n}}/{2t}}\label{prob},\ee
where $\mathcal{N}$ is a normalization constant.


\section{Dynamics of the resolvent in the large N limit}

In this section, we analyze the bulk properties of the spectrum of eigenvalues by deriving an equation for the average density from Eq.~(\ref{SFP3}). This is possible in the limit of large matrices. Since we are dealing with rectangular matrices, care must be taken in taking this limit. Let us define the rectangularity $r\equiv {N}/{M}$ and recall that we already set $\nu\equiv M-N$. The limit $M\to\infty$, $N\to\infty$ can be taken either at $r$ fixed, in which case $\nu\to\infty$, or at $\nu$ fixed, in which case $r\to 1$. Most of the time we shall work at fixed $r$. 

\subsection{The partial differential equation for the Greens function of the Wishart random martix}

We start by recalling some definitions. 
The averaged density of eigenvalues is defined by: 
\be\nonumber
\tilde{\rho}\left(\lambda,t\right)&=&\int\prod_{k=1}^{N}d\lambda_{k}\,P\left(\lambda_{1},\cdots,\lambda_{N}\right)\sum_{l=1}^{N}\delta\left(\lambda-\lambda_{l}\right)=\nonumber\\
&=&\left\langle \sum_{l=1}^{N}\delta\left(\lambda-\lambda_{l}\right)\right\rangle.\ee
One defines similarly the `two-particle' density  \be \tilde{\rho}\left(\lambda,\mu,t\right)=\left\langle \sum_{l=1}^{N}\sum_{j\left(\neq l\right)}\delta\left(\lambda-\lambda_{l}\right)\delta\left(\mu-\lambda_{j}\right)\right\rangle.\ee
These densities are normalized as follows
\be
\int d\lambda\,\tilde \rho(\lambda,t)=N,\quad 
\int d\lambda d\mu \,\tilde\rho(\lambda,\mu,t)=N(N-1).
\ee
Furthermore, observe  that \be \sum_{j\left(\neq i\right)}\frac{1}{\lambda_{i}-\lambda_{j}}=\dashint\frac{1}{\lambda_{i}-\mu}\sum_{j\left(\neq i\right)}\delta\left(\lambda_{j}-\mu\right)d\mu,\label{trick1}\ee
with $\dashint$ denoting the principal value of the integral.

The equation for $\tilde{\rho}\left(\lambda,t\right)$ is easily obtained by multiplying Eq.~ (\ref{SFP3}) by $\sum_{l=1}^{N}\delta\left(\lambda-\lambda_{l}\right)$ and integrating over all $N$ eigenvalues. One gets\be\nonumber
\frac{\del\tilde{\rho}\left(\lambda,t\right)}{\del t}=2\frac{\del^{2}}{\del\lambda^{2}}\left[\lambda\tilde{\rho}\left(\lambda,t\right)\right]
+\beta\left[N\left(1-\frac{1}{r}\right)-1\right]\\
\times\frac{\del}{\del\lambda}\tilde{\rho}\left(\lambda,t\right)
-2\beta\frac{\del}{\del\lambda}\left[\lambda\dashint\frac{\tilde{\rho}\left(\lambda,\mu,t\right)}{\lambda-\mu}d\mu\right].\label{densities1}
\ee
This equation simplifies in the limit of very large matrices. To see that, let us set $\tilde{\rho}\left(\lambda,\mu\right)=\tilde{\rho}\left(\lambda\right)\tilde{\rho}\left(\mu\right)+\tilde{\rho}_{con}\left(\lambda,\mu\right)$, where $\tilde{\rho}_{con}\left(\lambda,\mu\right)$ is the connected part of the two-point density, expected to be of order $1/N$ as compared to the factorized contribution $\tilde{\rho}\left(\lambda\right)\tilde{\rho}\left(\mu\right)$. Next, we simultaneously rescale the time, so that $\tau=Nt$, and change the normalization of the densities: 
\be\tilde{\rho}\left(\lambda\right)=N\rho\left(\lambda\right),\quad \tilde{\rho}\left(\lambda,\mu\right)=N\left(N-1\right)\rho\left(\lambda,\mu\right).\ee
This results in the following equation 
\be\nonumber
&&\frac{\del\rho\left(\lambda,\tau\right)}{\del\tau}+\beta\left(\frac{1}{r}-1\right)\frac{\del\rho\left(\lambda,\tau\right)}{\del\lambda}+\nonumber\\
&&+2\beta\frac{\del}{\del\lambda}\left[\lambda\rho\left(\lambda,\tau\right)\dashint\frac{\rho\left(\mu,\tau\right)}{\lambda-\mu}\right]=\frac{2}{N}\lambda\frac{\del^{2}\rho\left(\lambda,\tau\right)}{\del\lambda^{2}}+\nonumber\\
&&+\frac{4-\beta}{N}\frac{\del\rho\left(\lambda,\tau\right)}{\del\lambda}-2\beta\frac{\del}{\del\lambda}\left[\lambda\dashint\frac{\rho_{con}\left(\lambda,\mu,\tau\right)}{\lambda-\mu}d\mu\right],\label{densities2}\ee
 where on the right side of the equal sign, we have gathered terms that disappear in the limit $N,M\to\infty$ with $r $ fixed. 
 
 At this point it is convenient to introduce the usual resolvent (sometimes also called Green's function)
\be 
G\left(z,\tau\right
)=\frac{1}{N} \left<\Tr \frac{1}{z-L(\tau)}\right>=\int d\mu\,
\frac{\rho(\mu, \tau)}{z-\mu}.
\ee
The  imaginary part of  $G(z=\lambda-i\epsilon,\tau)$ yields the average spectral density, whereas the real part is the Hilbert transform of $\rho$ ($\mathcal{H}\left[\rho\left(\lambda\right)\right]=\dashint\frac{\rho\left(\mu\right)}{\lambda-\mu}d\mu$). In the limit  $N,M\to\infty$, with $r$ fixed,  we then obtain the following  closed equation for $\rho\left(\lambda,\tau\right)$:
\be\nonumber
&&\frac{\del\rho\left(\lambda,\tau\right)}{\del\tau}=\beta\left(1-\frac{1}{r}\right)\frac{\del\rho\left(\lambda,\tau\right)}{\del\lambda}+\\
&&-2\beta\lambda\frac{\del}{\del\lambda}\left\{\rho\left(\lambda,\tau\right)\mathcal{H}\left[\rho\left(\lambda,\tau\right)\right]\right\}-2\beta\rho\left(\lambda,\tau\right)\mathcal{H}\left[\rho\left(\lambda,\tau\right)\right].\nonumber\\\label{densities3}\ee
One can take its Hilbert transform to obtain: 
\be
&&\frac{\del\mathcal{H}\left[\rho\left(\lambda,\tau\right)\right]}{\del\tau}=
\beta\left(1-\frac{1}{r}\right)\frac{\del\mathcal{H}\left[\rho\left(\lambda,\tau\right)\right]}{\del\lambda}\nonumber\\
&&+2\beta\lambda\rho\left(\lambda,\tau\right)
\frac{\del\rho\left(\lambda,\tau\right)}{\del\lambda}
-2\beta\lambda\mathcal{H}\left[\rho\left(\lambda,\tau\right)\right]\frac{\del\mathcal{H}\left[\rho\left(\lambda,\tau\right)\right]}{\del\lambda}
\nonumber\\&&+\beta\left[\rho\left(\lambda,\tau\right)\right]^{2}-\beta\left\{\mathcal{H}\left[\rho\left(\lambda,\tau\right)
\right]\right\}^{2}.\label{Hdensities}\ee
where we have used that for  $f(x)=\frac{{\rm d}}{{\rm d}x}\left\{\rho(x)\mathcal{H}\left[\rho(x)\right]\right\}$, $\mathcal{H}\left[xf(x)\right]=x \mathcal{H}\left[f(x)\right]$.
By combining the two equations above, we derive the equation for the resolvent that is the  analogue of the Burgers equation obtained earlier ~\cite{BN2, NEUB2} for the GUE and CUE ensembles:
\be \partial_{\tau}G(z,\tau)=\beta(1\!-\!\frac{1}{r}\!-\!2zG(z,\tau))\partial_z G(z,\tau)\!-\!\beta G^2(z,\tau).
\label{wishartburgers1} \ee 
This equation has been obtained in a slightly different form in Ref.~\cite{DUGU}, using other techniques. To establish the correspondence with the equation derived in Ref.~\cite{DUGU}, we  need to rescale $\tau\to\frac{r\tau}{\beta}$. Then Eq.~(\ref{wishartburgers1}) transforms into \be\nonumber
&&\partial_{\tau}G(z,\tau)=-\partial_z G(z,\tau)\nonumber\\
&&+r\left(\partial_z G(z,\tau)-2zG(z,\tau)\partial_z G(z,\tau)-G^2(z,\tau)\right), \label{wishartburgers2}
\ee
in full agreement with \cite{DUGU}. This nonlinear partial differential equation, governing the evolution of the resolvent of a freely diffusing Wishart matrix, will be solved in the next subsection using the method of (complex) characteristics \cite{DUGU}, as we did in our earlier works \cite{BN1,BN2}. 

\subsection{Solution with complex characteristics}

It is convenient to rewrite Eq.~(\ref{wishartburgers2}) obeyed by $G$ as follows, 
\be	\left({1-r+2r z G }\right)\frac {\partial G}{\partial z }+\frac {\partial G}{\partial \tau}+r{G}^{{2}}=0, \ee
from which the equations for the characteristic lines (parameterized by $s$)  are easily obtained
\be\frac {{\rm d}z}{{\rm d}s}=1-r+2rzG,\label{ode3}\ee
\be\frac {{\rm d}\tau}{{\rm d}s}=1\label{ode1},\ee
	\be\frac {{\rm d}G}{{\rm d}s}=-r{G}^{{2}}\label{ode2}.\ee
We choose to solve these equations with the same initial condition that leads to Eq.~(\ref{prob}), namely $G\left({z,{\tau =0}}\right)={1}/{z}$. This  translates into $z(s=0)=z_0, \tau(s=0)=0,G(s=0)={1}/{z_{0}}$.  Solving the last  two equations gives then  respectively $s=\tau$ and 
\be G=\frac{1}{r\tau+z_{0}}.\label{G1}\ee 
We are therefore left with:
\be
\frac{{\rm d}z}{{\rm d}\tau}=1-r+\frac{2rz }{r\tau+z_{{0}}},
\ee
whose solution reads:
	\be z=\left({1+\frac {\tau}{{z}_{{0}}}}\right)\left({{z}_{{0}}+r\tau}\right).\label{char1}\ee
One can then use this result to eliminate $z_0$ in Eq.~(\ref{G1}), which yields the following implicit equation for $G(z,\tau)$:	
	\be \label{Gimplicit} z=\frac{1}{G(z,\tau)}+\frac{\tau}{1-r\tau G(z,\tau)}.\ee
	
The construction just outlined supposes that the mapping between $z$ and $z_0$ is one-to-one, that is, it can be inverted. This is the case except at points where ${\rm d}z/{\rm d}z_0=0$ where a singularity occurs, that may be interpreted as the formation of a ``preshock'' in the Burgers equation. The singular points are determined by 
\be
\frac{dz}{dz_0}=1-\frac{r}{z_0^2}\tau=0 ,\qquad z_{0c}=\pm \sqrt{r}\tau,
\ee
corresponding to $z_c=(1\pm\sqrt{r})^2 \tau$. These singularities occur  precisely at the edges of the spectrum, as we shall verify shortly.  Note that the left singularity (for positive $z$) originates from characteristics starting at negative $z_0$.   

The solution of Eq.~(\ref{Gimplicit}) reads explicitly
\be  G(z,\tau)=\frac{(r\!-\!1)\tau+z-\sqrt{z^2\!-\!2z\tau(1\!+\!r)\!+\!\tau^2(1\!-\!r)^2}}{2r\tau z},\label{wishartgreen}\ee
where the (minus) sign in front of the square root is chosen so that the resulting spectrum density is positive. 
$G(z,\tau)$ matches its initial condition when $\tau\to 0$. We have also that $G(z, \tau)$ behaves as $1/z$ at large $z$ and fixed $\tau$, as appropriate since  the spectrum is bounded. By taking the imaginary part of $G(z=\lambda-i\epsilon, \tau)$, one recovers the  well known Bronk-Marcenko-Pastur formula for the level density:
\be
	\rho \left({\lambda ,{\tau }}\right)=\frac {\sqrt{{\left({\lambda -{c}_{{-}}\tau }\right)\left({{c}_{{+}}\tau -\lambda }\right)}}}{2\pi \lambda \tau r}.\label{BMP}
\ee
The spectrum is  localized in the interval $[c_{-}\tau,c_{+}\tau]$ with $c_{\pm}=(1\pm\sqrt{r})^2$, as anticipated from the study of the preshocks.

\subsection{The partial differential equations for related ensembles}
A similar treatment can be applied for the ensembles closely related to the Wishart ensemble. For instance, the resolvent  for the chiral matrix  is defined as
\be
g(w,\tau)\equiv\frac{1}{N+M} \left\langle\Tr\frac{1}{w-W(\tau)}\right\rangle
\ee
It is connected to its Wishart equivalent by the following transformation (see e.g. ~\cite{FEINZEE}):
\be
	G\left({z}\right)=\frac {r-1}{2r{w}^{{2}}}+\frac {r+1}{2rw}g\left({w}\right),\label{wisharttochiral}
\ee
where $w^{2}=z$. By inserting this change of variable into Eq.~(\ref{wishartburgers2}), one gets
\be
	{{\left(\frac{1-r}{1+r}\right)}}^{{2}}+{w}^{{3}}\left[{\frac{2}{1+r}\partial_{\tau} g+g\partial_w g}\right]=0.\label{chiralburg}
\ee
For $M=N$, i.e., $r=1$, this equation reduces to the same Burgers equation that was obtained for the diffusing Hermitian matrix in \cite{BN2}.

We can additionally relate the Green's function of the Wishart ensemble to that representing random anti-Wishart matrices. We have  ~\cite{FEINZEE,MODERNWISHART}:  
\be
G_{a}(z)\equiv \frac{1}{M} \left\langle\Tr\frac{1}{z-L_{a}}\right\rangle=\frac{1-r}{z}+rG(z),
\ee
a relation which also holds in the case when the matrices evolve in time (the  term $1/z$ corresponds to the zero modes of the matrix, which are of `kinematical' origin,  reflecting the rank of the matrix).   The analog of Eq.~(\ref{wishartburgers2})  for $G_a$ is 
\be
 \partial_{\tau}G_{a}(z,\tau)=\left(1\!-\!r\!-\!2zG_{a}(z,\tau)\right)\partial_z G_{a}(z,\tau)\!-\!G_{a}^2(z,\tau). \label{Awishartburgers2}
\ee

The Green's function for the chiral matrix  can be easily obtained from (\ref{wishartgreen}) by using the relation  (\ref{wisharttochiral}) between $G(z,\tau)$ and $g(w,\tau)$. One gets
\be
g\left(w,\tau\right)=\frac{w^2-\sqrt{\left(w^2\!-\!c_{-} \tau\right) \left(w^2\!-\!c_{+}\tau\right)}}{(r+1) \tau w}.\label{chiralgreen}
\ee
Similarly, the anti-Wishart Greens function reads
\be
G_{a}(z,\tau)=\frac{(1-\!r)\tau\!+\!z\!-\!\sqrt{z^2\!-\!2z\tau(1\!+\!r)+\tau^2(1\!-\!r)^2}}{2\tau z}\label{Asol1}.
\ee
Note the pole at $z=0$, which reflects the presence of the zero modes that we have already mentioned. 
The residue, which reads $1-r=\frac{M-N}{M}$, accounts exactly for the ratio of the number of zero modes to the total number of eigenvalues.

\subsection{Connection to the Voiculescu approach}

It is interesting to observe that Eq.~(\ref{Gimplicit}) that emerged from analyzing the complex characteristics of Eq.~(\ref{wishartburgers2}) can also be obtained from the free random variable formalism \cite{VOICULESCU}. The cornerstone of this approach is the so-called R-transform, which plays the role of the logarithm of the characteristic function in the classical probability calculus. The R-transform is additive under free convolution, alike the logarithm of the characteristic function, generating additive cumulants for the convolution of independent probability distributions.  The R-transform is related to the Green's function by  
\be R[G(z)]+1/G(z)=z, 
\label{Rdef}
\ee
or equivalently  $G[R(z)+1/z]=z$. Thus, modulo the shift $1/z$, the R-transform is the functional inverse of the Green's function.  

The R-transform  for a static Wishart random matrix is known (see e.g. \cite{PETZ}). It reads $R_{st}(z)=1/(1-rz)$. 
In order to get the R-transform of the time-dependent Wishart matrix $L(\tau)$, we note that the  time evolution of the probability distribution $P(\tau)$ is equivalent to a linear rescaling  with time of the eigenvalues in a stationary probability distribution function $P_{st}$ (see e.g. Eq.~(\ref{prob})). We may therefore write
\be
G(z,\tau)&=&\frac{1}{N}\left\langle\Tr\frac{1}{z-L(\tau)}\right\rangle_{P(\tau)}\nonumber\\
&=&\frac{1}{N}\left\langle\Tr\frac{1}{z-\tau L_{st}}\right\rangle_{P_{st}}=\frac{1}{\tau}G_{st}(\frac{z}{\tau}).
\ee
Then,  by using the definition of the R-transform, Eq.~(\ref{Rdef}), and the explicit form of the R-tranform of the Wishart matrix recalled above, we get  $R(z,\tau) =\tau R_{st} (\tau z)$. We see therefore that the quadratic equation~(\ref{Gimplicit})  emerging as the solution of the complex characteristics equation is simply Eq.~(\ref{Rdef})  for 
the time-dilated R-transform  for the Wishart ensemble.

\section{Exact equation for the average characteristic polynomial and its scaling limits}

As it has been already emphasized in \cite{BN2}, it does not appear possible to push the simple, and physically intuitive, description of the level density given in the previous section, beyond the large $N$ limit, since the $1/N$ corrections are difficult to calculate. In order to study these finite size corrections, we shall then proceed as in \cite{BN2}, and use the technique of orthogonal polynomials. After a suitable generalization of the relevant polynomials to the time-dependent problem, we obtain for these polynomials exact equations. In particular the logarithm of the average characteristic polynomial is found to obey an exact equation, which in the  large $N$ limit coincides with the equation for the resolvent that was discussed in the previous section. This allows us to study universal behavior in the vicinities of the edges of the (large $N$) spectrum, albeit not of the spectral density itself but of the logarithm of the average characteristic polynomial.  
\subsection{Time dependent characteristic polynomials}
 For the chosen initial conditions, the average characteristic polynomial associated with the diffusing Wishart matrix is equal to a certain time dependent monic orthogonal polynomial \cite{BHPOLS}
\be
\left\langle\det\left[z-L(\tau)\right]\right\rangle=M_{N}(z,\tau).
\ee
In the  static case the relevant  orthogonal polynomials are the generalized Laguerre polynomials \cite{FYOSTRA}. These are defined as
\be 
L^{\alpha}_{n}(x)=\sum^{n}_{j=0}\frac{(-x)^{j}}{j!}\left(\begin{array}{c}n+\alpha \\ n-j \end{array}\right)\label{Lpoly}, 
\ee  
and satisfy the orthogonality relation 
\be\int^{\infty}_{0} e^{-x}x^{\alpha}L^{\alpha}_{n}(x)L^{\alpha}_{m}(x)=\delta_{nm}\frac{\Gamma(n+\alpha+1)}{n!} \label{Lorth}.\ee
The following recursion relation
\be n L^{\alpha}_{n}(x)=(\alpha+1-x)L^{\alpha+1}_{n-1}(x)-x L^{\alpha+2}_{n-2}(x)\label{Lrr}  ,\ee
and the differentiation property
\be \frac{{\rm d}^{k}}{{\rm d}x^{k}}L^{\alpha}_{n}(x)=(-1)^{k}L^{\alpha+k}_{n-k}(x) \label{Ldiff},\ee
will also be useful. 

For the dynamical case, we need to construct time dependent, monic, polynomials that are orthogonal with respect to the measure defined by the solution of the Smoluchowski-Fokker-Planck equation (\ref{prob}) for $\beta=2$. It is easy to check that the following polynomials satisfy these requirements
\be 
M^{\alpha}_{n}\left(x,\tau\right)=\left(-\tau\right)^{n}n! L^{\alpha}_{n}\left(\frac{x}{\tau}\right) \label{Mpoly},
\ee 
with the orthogonality condition given by
\be\nonumber
&&\int^{\infty}_{0} e^{-\frac{x}{\tau}}\left(\frac{x}{\tau}\right)^{\alpha}M^{\alpha}_{n}\left(x,\tau\right)M^{\alpha}_{m}\left(x,\tau\right)\nonumber\\
&&\qquad\qquad\qquad=\delta_{nm}\left(\tau\right)^{2n+1}\Gamma(n+\alpha+1)n!\label{Morth}\; .
\ee 
The analogues of Eqs.~(\ref{Lrr}) and (\ref{Ldiff}) read respectively
\be\nonumber
n M^{\alpha}_{n}\left(x,\tau\right)=n\left[x-\tau(\alpha+1)\right]M^{\alpha+1}_{n-1}\left(x,\tau\right)+\\
-n(n-1)\tau x M^{\alpha+2}_{n-2}\left(x,\tau\right), \label{Mrr}
\ee 
and 
\be 
\frac{\del^{k}}{\del x^{k}} M^{\alpha}_{n}\left(x,\tau\right)=\frac{n!}{(n-k)!}M^{\alpha+k}_{n-k}\left(x,\tau\right) \label{Mdiff}.
\ee

In order to derive the differential equation satisfied by the polynomials $M^{\alpha}_{n}\left(x,\tau\right)$ we first observe   that
\be \frac{\del}{\del\tau} M^{\alpha}_{n}(x,\tau)=\frac{n}{\tau} M^{\alpha}_{n}(x,\tau)-\frac{x}{\tau}\frac{\del}{\del x} M^{\alpha}_{n}(x,\tau) \label{delTauM}.\ee
Then, using  (\ref{Mrr}), we eliminate the first term on the right hand side of this equation, which yields:
\be \del_{\tau}M^{\alpha}_{n}(x,\tau)=-x\del_{xx}M^{\alpha}_{n}(x,\tau)-\left(1+\alpha\right)\del_{x}M^{\alpha}_{n}(x,\tau).\ee
At this point, we rescale the time in order to be  consistent with our earlier calculations, namely we set: $\tau\to\frac{2r\tau}{\beta N}$ (with $\beta=2$). We also set $\alpha=\nu$. Finally, since we are interested mostly in the average characteristic polynomial, we focus on $n=N$. We thus obtain the following exact partial differential equation for the characteristic polynomial:
\be \del_{\tau}M^{\nu}_{N}(z,\tau)\!=-\!\frac{r}{N}\left[z\del_{zz}M^{\nu}_{N}(z,\tau)\!+\!\left(1\!+\!\nu\right)\del_{z}M^{\nu}_{N}(z,\tau)\right].\label{Meq}\ee

It is easy to show that this equation yields Eq.~(\ref{wishartburgers2}) for the resolvent in the large $N$ limit. To do so, let us recall first that  $\del_{z}\ln\left\langle\det\left[z-L(\tau)\right]\right\rangle=NG(z,\tau)$ in the large $N$ limit. We then define 
\be
f_{N}(z,\tau)\equiv\frac{1}{N}\del_{z}{\rm ln}\left[M^{\nu}_{N}(z,\tau)\right]\label{deff}
\ee 
(the inverse Cole-Hopf transform), and obtain from Eq.~(\ref{Meq}) the following equation for $f_N$
\be\nonumber
&&\del_{\tau}f_{N}+r\left(2zf_{N}\del_{z}f_{N}+f_{N}^{2}\right)+(1-r)\del_{z}f_{N}\nonumber\\
&&=-\frac{r}{N}\left(2\del_{z}f_{N}+z\del_{zz}f_{N}\right).\label{eqf2}
\ee
In the large $N$ limit, the right hand side can be dropped, and one recovers Eq.~(\ref{wishartburgers2}) for  $f_{N}(z,\tau)=G(z,\tau)$.
  
The exact equation (\ref{Meq}) that we have obtained will be used in the following subsection in order to analyze the $1/N$ corrections of the characteristic polynomial in the vicinity of the edges of the spectrum.

\subsection{Characteristic polynomial at the edge of the spectrum}

We are interested in the asymptotic scaling of the averaged characteristic polynomial at the  edge of the spectrum, where the average eigenvalue spacing is of  order $N^{-\delta}$. The power $\delta$ is determined by the singular behavior  of the spectrum (in the large $N$ limit) near its edge. It is easily verified that a power law behavior $\sim |z-z_c|^\alpha$ yields $\delta=1/(\alpha+1)$. We shall then write $z=z_{c}(\tau)+ N^{-\delta}s$, and study the behavior of the characteristic polynomial, or its inverse Cole-Hopf transform, as a function of $s$ in the large $N$ limit. We shall set
\be
f_{N}\left(z_{c}(\tau)+ N^{-\delta}s,\tau\right)\approx A(\tau)+N^{-\gamma}\chi(s,\tau),\label{ansatz1}
\ee
where the function $\chi(s,\tau)$ remains finite in the large $N$ limit, and  $\gamma=1-\delta$ (for a singularity $\sim |z-z_c|^\alpha$,  $\gamma=\alpha/(\alpha+1)$).

Two distinct limiting behaviors arise: the first one, which  occurs at the right shock for any $r$ and at the left one for $r\ne 1$, is referred to as  `soft edge scaling'; the second, obtained at the left shock for $r=1$, i.e.  at the origin, is known as  `hard edge scaling'. We shall examine successively these two cases.

\subsubsection{The soft edge}
We know the solution $f_{\infty}(z,\tau)$ of Eq.~(\ref{eqf2}) in the limit  $N,M\to\infty$ and fixed $r$. This is given by Eq.~(\ref{wishartgreen}), i.e., 
\be f_{\infty}(z,\tau)=\frac{(r-1)\tau+z-\sqrt{(z-z_{L})(z-z_{R})}}{2r\tau z}, \label{soleqf}
\ee 
where $z_{L}=\tau(1-\sqrt{r})^{2}$ and $z_{R}=\tau(1+\sqrt{r})^{2}$ are the positions of the left and right  edges of the spectrum, respectively.
As announced above, we either probe the close vicinity of the left $z=z_{L}+N^{-\delta}s$ ($r\ne 1$) or the right $z=z_{R}+N^{-\delta}s$ traveling shocks. At those points the average eigenvalue spacing is proportional to $N^{-2/3}$ and so $\delta=\frac{2}{3}$, corresponding to a square root singularity ($\alpha=1/2$).  Our ansatz (\ref{ansatz1}) has therefore the following form:
\be f^{L/R}_{N}(z_{_{L/R}}+N^{-\frac{2}{3}} s,\tau)=\frac{(r-1)\tau+z_{_{L/R}}}{2r\tau z_{_{L/R}}}+N^{-\frac{1}{3}}\chi(s,\tau),\label{fcritA1}
\ee
Inserting it into (\ref{eqf2}) and keeping the dominant terms as $N\to\infty$, one obtains:
\be -\frac{1}{z_{*}}+2\chi\del_{s}\chi+\del_{ss}\chi=0,\label{eqAiry1}\ee 
where we define $z_{*}=\{-z^{2}_{L}r^{\frac{3}{2}}\tau,z^{2}_{R}r^{\frac{3}{2}}\tau\}$ depending on the edge we look at. This equation  can  be written 
\be \del_{s}\left[\chi^{2}+\del_{s}\chi-\frac{s}{z_{*}}\right]=0, \label{eqAiry2}\ee
which is easily integrated to yield 
\be \chi^{2}+\del_{s}\chi-\frac{s}{z_{*}}+g(\tau)=0,\label{eqAiry3}\ee
where $g(\tau)$ is an arbitrary  function of $\tau$. In terms of $\phi(s,\tau)$ defined as 
$\chi(s,\tau)=\partial_s \ln \phi(s,\tau)$ (inverse Cole-Hopf transform) one obtains
\be \del_{ss}\phi(s,\tau)+\left(g(\tau)-\frac{s}{z_{*}}\right)\phi(s,\tau)=0.\label{eqAiry4}\ee
The shift of variable $s=y+g(\tau)z_{*}$,  and the redefinition $\psi(y)=\phi(y+g(\tau)z_{*})$,  transform this equation into the  equation for the Airy function
\be \del_{yy}\psi(y,\tau)-\frac{y}{z_{*}}\psi(y,\tau)=0, 
\ee
The solution reads:
\be \phi(s,\tau)= {\rm Ai}\left(\frac{s-g(\tau)z_{*}}{\sqrt[3]{z_{*}}}\right).\label{solAiry1}\ee
In order to find the unknown function $g(\tau)$ we match the large $s$ asymptotics of (\ref{fcritA1}) with the Green's function (\ref{wishartgreen})  at $|z|\to z_{_{L/R}}$.  This way (see the Appendix) we find that $g(\tau)=0$.

\subsubsection{The hard edge}
The spectrum of the infinite size $L(\tau)$ matrix, for any $\tau>0$, ``touches" the origin only if  the rectangularity $r\simeq 1$. This is therefore when we can observe the hard edge universal scaling. To this end we rewrite (\ref{eqf2}) in the following form:
\be \del_{\tau}f_{N}\!+\! r\left(2zf_{N}\del_{z}f_{N}\!+\!f_{N}^{2}\right)\!=\!-\!\frac{r}{N}\left[(2\!+\! \nu)\del_{z}f_{N}\!+\! z\del_{zz}f_{N}\right].\label{eqf4}\ee
In the limit of $M$  and $N$ going to infinity with $\nu$ fixed and finite (so that $r\to 1$), we have:
\be f_{\infty}(z,\tau)=\frac{z-\sqrt{z^{2}-4\tau z}}{2\tau z}.
\label{greenzero}\ee
In  the vicinity of the origin, the singularity is of the inverse square root type, so that the average eigenvalue spacing is proportional to $N^{-2}$, or  $\delta=2$. We set then $z=N^{-2}s$ and Eq.~(\ref{ansatz1}) takes the following form:  
\be f_{N}(N^{-2}s,\tau)=\frac{1}{2\tau}+N \chi(s,\tau).\label{fN1}\ee
Inserting (\ref{fN1}) into (\ref{eqf4}) and taking the large $N$ limit with $r\to 1$ and $\nu$ fixed, one obtains:
\be s\del_{ss}\chi+2s\chi\del_{s}\chi+(2+\nu)\,\del_{s}\chi+\chi^{2}=0.\label{eqBessel1}\ee
We now have
\be
\del_{s}\left[s\del_{s}\chi+s\chi^{2}+(1+\nu)\chi\right]=0 \label{chi2}
\ee
and therefore
\be 
s\del_{s}\chi+s\chi^{2}+(1+\nu)\chi+g(\tau)=0 \label{chi3},
\ee
$g(\tau)$ being an unknown function of $\tau$. For $\phi(s,\tau)$, this implies:
\be s\del_{ss}\phi+(1+\nu)\del_{s}\phi+g(\tau)\phi=0\label{BesseleClliford}.\ee
We now proceed by setting $s=\left[h(\tau)y\right]^{2}$, with $h(\tau)$ an arbitrary function of $\tau$, and also $\phi(s)=\left[h(\tau)y\right]^{-\nu}\psi(y)$. In terms of $\psi(y)$, Eq.~(\ref{BesseleClliford}) takes the following form:
\be y^{2}\del_{yy}\psi+y\del_{y}\psi+\left[4g(\tau)h^{2}(\tau)y^{2}-\nu^{2}\right]\psi=0\label{psi1}.\ee 
If we now choose  $h(\tau)=\frac{1}{2\sqrt{g(\tau)}}$, we obtain:
\be y^{2}\del_{yy}\psi+y\del_{y}\psi+\left(y^{2}-\nu^{2}\right)\psi=0\label{psi2},\ee
which we recognize as the Bessel equation. The relevant solutions are
$\psi(y)={\rm J}_{\nu}(y)$, 
and therefore
\be \phi(s)=s^{-\frac{\nu}{2}}{\rm J}_{\nu}\left(2\sqrt{g(\tau)s}\right)\ee
Alike in the case of the soft age, we again match the asymptotic behavior of (\ref{fN1}) for $|s| \rightarrow \infty$  with the Green's function (\ref{greenzero}) at $|z|\to 0$.  In this way (see the Appendix) we  get  that $g(\tau)=\frac{1}{\tau}$.
\subsection{Equation for the Cauchy transform of the orthogonal polynomials}

Following our previous paper \cite{BN2}, we conclude by briefly describing the evolution of the average of the inverse of the characteristic polynomial. In the case of our specific initial conditions (see after Eq.~(\ref{SFP3})), it is proportional  to the following Cauchy transform of the time-dependent monic orthogonal polynomial:
\be
p^{\nu}_{N-1}(z,\tau)=\frac{1}{2\pi i}\int_{0}^{\infty}\dd x\frac{M^{\nu}_{N-1}(z,
\tau)\left(\frac{x}{\tau}\right)^{\nu}\exp{\left(-\frac{rx}{(N-1)\tau}\right)}}{x-z}.\nonumber\\
\ee
A straightforward (albeit tedious) calculation reveals that, given (\ref{Meq}), the object defined by
\be
\tilde{p}^{\nu}_{N}(z,\tau)\equiv \left(\frac{\tau}{z}\right)^{\nu}\exp{\left(\frac{rz}{N\tau}\right)} \,p^{\nu}_{N}(z,\tau)
\ee
satisfies:
\be 
\del_{\tau}\tilde{p}^{\nu}_{N}(z,\tau)=-\frac{r}{N}\left[z\del_{zz}\tilde{p}^{\nu}_{N}(z,\tau)+\left(1+\nu\right)\del_{z}\tilde{p}^{\nu}_{N}(z,\tau)\right],\label{Pteq}
\ee
a partial differential equation of exactly the same form as Eq.~(\ref{Meq}). The difference resides in the associated initial conditions which, for the Cauchy transforms are singular at the origin.  This results in a different choice of the corresponding solutions. In the case of the soft edge we recover the GUE result~\cite{BN2}, i.e. Airy functions of the second kind. In the case of the hard edge, one gets Bessel functions of the third  kind (Hankel functions), in agreement with Ref.~\cite{AF} for the respective universality classes.

\section{Conclusions}
In this paper,  we have studied a freely diffusing matrix of the Wishart ensemble. The  Smoluchowski-Fokker-Planck equation for the stochastic evolution of its eigenvalues allowed us to construct, after a proper time rescaling, two partial differential equations. The first equation, which has the form of a generalized Burgers equation,  describes the behavior of the associated Green's function in the limit of a matrix of infinite size. We have solved this equation with the method of complex characteristics for the particular case where initially the matrix vanishes. We related the singularities of the characterstics to the occurrence of pre-shocks in the Burgers equation. Additionally, we translated this results into those describing chiral and anti-Wishart matrices and made connections to the method of free random variables. The second equation concerns the averaged characteristic polynomial which, for the chosen initial conditions, is equal to a certain time-dependent monic orthogonal polynomial of the Laguerre type. This partial differential equation, is an exact equation, for any matrix size, and it reduces to the generalized Burgers equation that has been obtained in the limit of infinite sizes. This equation lends itself to a scaling  analysis in the vicinity of the edges of the large $N$ spectrum. We then recovered the asymptotic scaling functions describing the universal pre-shocs at the ``soft edge", of the Airy type and, for the first time within the present approach, for the ``hard-edge" singularity - of the Bessel type. This provides further evidence of the  efficiency of the method. In a forthcoming publication~\cite{BNW4}, we shall present a generalization of the derivation of the equation for the average characteristic polynomial that does not rely on the use of orthogonal polynomials. This  will allow us to explore other types of initial conditions leading to a new type of  singularity.   In particular, we derive for Wishart and chiral complex ensembles a new type of cusp singularity (generalized Bessoid), which supersedes the known Pearcey cusp~\cite{BREZINHIKAMIGAP}  at the closure of the gap.

\section*{Acknowledgments}
PW would like to thank the organizers of the 2011 Les Houches school on vicious walkers and random matrices, where part of this work has been done.  
PW is  supported by the International PhD Projects Programme
of the Foundation for Polish Science within
the European Regional Development Fund of the European
Union, agreement no. MPD/2009/6. MAN is  supported in part by the Grant DEC-2011/02/A/ST1/00119 of the National Centre of Science. 

\section*{Appendix}

We start from matching the Airy edge. For $|z|\to\infty$ and $0 \le\arg(z)<\pi$, in the leading order~\cite{AS}\begin{equation}
	\frac{Ai'(z)}{Ai(z)}\sim-\sqrt{z},
	\end{equation}	
From (\ref{fcritA1}) 	we get
	\begin{eqnarray}
		\lim_{s\to\infty}&&f_{N}^{L/R}(z_{L/R}(\tau)+N^{-2/3}s, \tau) \approx A(\tau)+ \lim_{s\to\infty}N^{-1/3}\frac{\partial_{s}Ai(x(s))}{Ai(x(s))} \nonumber \\
		&=&A(\tau)+\lim_{x\to\infty}N^{-1/3} z_{*}^{-1/3} \frac{Ai'(x)}{Ai(x)} \nonumber \\
		&=&A(\tau)- z_{*}^{-1/2}\sqrt{s-g(\tau)z_{*}}\, .
	\end{eqnarray}
	where for simplicity we denoted the first term of the r.h.s. of (\ref{fcritA1}) as $A(\tau)$.
	This expression has to match (for any argument of $s$ except $\pi$) the  corresponding limit for the resolvent, i.e. 
	\begin{equation}
		\lim_{z\to z_{_{L/R}}} G(z)=A(\tau)-\lim_{z\to z_{_{L/R}}}\frac{\sqrt{(z-z_{_{L}})(z-z_{_{R}})}}{2r\tau z},
	\end{equation}
	with $z=z_{_{L/R}}+N^{-2/3}s$. 
	An elementary calculation shows  that
	\begin{equation}
	-\frac{\sqrt{(z_{_{R}}-z_{_{L}})(-s)}}{2r\tau z_{_{L}}}=-\sqrt{\frac{s}{-z_{_{L}}^{2}r^{3/2}\tau}}
	\end{equation}
	and
	\begin{equation}
	-\frac{\sqrt{(z_{_{R}}-z_{_{L}})s}}{2r\tau z_{_{R}}}=-\sqrt{\frac{s}{z_{_{R}}^{2}r^{3/2}\tau}},
	\end{equation}
	for the two edges, respectively,  so that the above equalities impose the condition $g(\tau)=0$.

For the matching of the Bessel edge, let us note~\cite{AS}  that 	
\begin{equation}
		\frac{J_{\nu}'(x)}{J_{\nu}(x)}\sim -i\, ,
	\end{equation}
 for $|x| \rightarrow \infty$ and  $0<\arg(x)<\pi$. 
For the Bessel edge we have 
\begin{equation}
	\phi(s,\tau)=s^{-\nu/2}J_{\nu}(2\sqrt{g(\tau)s}).
\end{equation}
Therefore, for $s\to\infty$ and $0< \arg(s)<2\pi$, a similar calculation as the one above yields, asymptotically, 
\begin{equation}
	f_N(N^{-2}s, \tau)= \frac{1}{2\tau} - i N \frac{\sqrt{g(\tau)}}{{\sqrt{s}}} .
\end{equation}
From the resolvent, we get
\begin{equation}
	\lim_{z=N^{-2}s \to 0^{+}}G(z)=\frac{1}{2\tau}-i\frac{1}{\sqrt{\tau z}}=\frac{1}{2\tau}-iN\frac{1}{\sqrt{\tau s}}
\end{equation}
which gives $g(\tau)=\frac{1}{\tau}$.


\end{document}